\documentclass[12pt]{iopart}
\usepackage{graphicx}
\begin{document}

\title{Geometrical structure effect on localization length of carbon nanotubes}
\author{Wen-Qi Ran$^{1}$\footnote{wenqi.ran@gmail.com}
, Jun Chang$^2$, Han-Tao Lu$^1$, Yue-Hua Su$^3$,
Hong-Gang Luo$^2$ and Tao Xiang$^2$}

\address{$^1$School of Physics, Peking University, Beijing 100871, China}

\address{$^2$Institute of Theoretical Physics and Interdisciplinary Center
of Theoretical Studies, Chinese Academy of Sciences, P.O. Box 2735, Beijing 100080,
China}

\address{$^3$Center for Advanced Study, Tsinghua University, Beijing 100084,
China}


\begin{abstract}
The localization length and density of states of carbon nanotubes are evaluated within
the tight-binding approximation. By comparison with the corresponding results for the
square lattice tubes, it is found that the hexagonal structure affects strongly the
behaviors of the density of states and localization lengths of carbon nanotubes.
\end{abstract}

\pacs{73.63.Fg, 61.43.Bn, 73.61.Wp}


\maketitle

In recent years, there have been extensive interests in the study of carbon nanotubes'
transport properties. \cite{1,2,3,4} In particular, extremely high or even ballistic
conductance was reported in carbon nanotubes. \cite{5,6} This indicates that carbon
nanotubes can be synthesized with high purity. However, as there are always some defects
or dislocations inside the tubes or caused by the substrate or attached to the tubes,
much attention has been paid to the study of the disorder effect in carbon nanotubes.
\cite{7,8} In 1992, Minitmire \cite{9} predicted that a carbon nanotube can behave as a
metal or a semiconductor depending on its chirality. This prediction was later confirmed
by scanning tunneling microscopic measurements. \cite{10} White argued that due to the
$C_{N}$ rotation symmetry the scattering of electrons is significantly reduced by the
doughnut-like wave packet confined along the tube but extended around its circumference
in the $(N,N)$ tubes. \cite{11} Furthermore, it is believed that semiconducting tubes are
more sensitive to long-range disorder than metallic tubes. \cite{12,13}

In this letter we explore the effect of geometrical structure on the scattering of
electrons in carbon nanotubes. We evaluate the localization length as well as the density
of states with the Green's function method. By comparison with the corresponding
quantities for the square lattice tubes, we find that the hexagonal structure can
significantly reduce the scattering of electrons by on-site random potentials. The
influence of the density of states (DOS) on the localization lengths of carbon nanotubes is also discussed.

Let us start with the Anderson model of random potentials for carbon nanotubes \cite{14}
\begin{equation}
H=\sum_{i}\varepsilon _{i}| i\rangle \langle i| +\sum_{\langle
ij\rangle}t | i\rangle \langle j |,
\end{equation}
where $\langle ij\rangle $ means that $i$ and $j$ are nearest
neighbors; $\varepsilon _{i}$ is a random on-site potential. It
takes any value between $-W$ and $W$ with equal probability.
Without loss of generality, we set $t=1$.

Carbon nanotubes can generally be classified by two integers and labeled as $(m,n)$. Two
types of carbon nanotubes are of particular interest. One is the armchair type with
$m=n$, and the other is the zigzag type with $n=0$. An undoped armchair nanotube is
always a metal, but a zigzag nanotube is a metal only when $m=3k$ with $k$ an integer. In
general, it shows that a carbon nanotube is a metal if $m-n$ is a multiple of $3$ or a
semiconductor otherwise. For simplicity, only the armchair or zigzag nanotubes will be
considered below.

\begin{figure}[!ht]
\centering \resizebox{3cm}{!} {\includegraphics{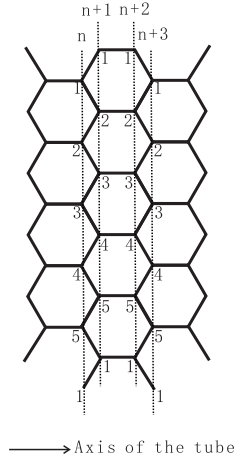}} \caption{The structure of a
unwrapped zigzag (5,0) nanotube. Each slice perpendicular to the tube axis contains five
sites connected by the dotted lines. } \label{Fig-structure}
\end{figure}

A carbon nanotube can be taken as a one dimensional stripe extending along the tube
direction. The width of the stripe depends on the structure of the nanotube. For a
$(m,0)$ zigzag tube, the width is $m$. If we use $|l\rangle$ to represent the basis state
for the $l$th slice of the stripe along the tube direction, then the Schr$\mathrm{\ddot{o}}$dinger
equation for the single-particle eigenfunction, $|\psi\rangle =\sum_{l} a_l |l\rangle $,
can be determined by the following recursion formula,
\begin{equation}
V_{l,l+1}a_{l+1}=\left( E-H_l\right) a_{l}-V_{l,l-1}a_{l-1},
\end{equation}
where $V_{l,l+1}$ is the interaction matrix between two
neighboring slices. $H_l$ is a diagonal matrix, formed by the
random on-site potential $\varepsilon_i$ in the $l$th slice. For
square lattices, $V_{l,l+1}$ is simply a unit matrix. However, for
carbon nanotubes, it takes a more complicated form. For a zigzag
tube, $V_{l,l+1}$ is a simple periodic function of $l$ and the
periodicity is 4. For example, for the $(5,0)$ zigzag nanotube
(Fig. 1), $V_{l,l+1}$ are

\begin{center}
\begin{eqnarray}
V_{4l,4l+1}&=& \left(
\begin{array}{ccccc}
1 & 1 & 0 & 0 & 0 \\
0 & 1 & 1 & 0 & 0 \\
0 & 0 & 1 & 1 & 0 \\
0 & 0 & 0 & 1 & 1 \\
1 & 0 & 0 & 0 & 1
\end{array}
\right),\nonumber \\
V_{4l+1,4l+2}&=&V_{4l+3,4l+4}= I_{5\times 5},\nonumber\\
V_{4l+2,4l+3}&=&\left(
\begin{array}{ccccc}
1 & 0 & 0 & 0 & 1 \\
1 & 1 & 0 & 0 & 0 \\
0 & 1 & 1 & 0 & 0 \\
0 & 0 & 1 & 1 & 0 \\
0 & 0 & 0 & 1 & 1
\end{array}
\right) ,\nonumber\\
\end{eqnarray}
\end{center}

where $I_{5\times 5}$ is a $5\times 5$ unit matrix. For other
zigzag or armchair tubes, $V_{l,l+1}$ can be similarly defined.

The localization length $\lambda$ is a characteristic length scale for describing the
decay of an eigenfunction in space. It is defined by \cite{15}
\begin{equation}
\frac{1}{\lambda }=-\lim_{l\rightarrow \infty }\frac{1}{2(l-1)}\ln
Tr\left\vert \left\langle 1\left\vert G\left( l\right) \right\vert
l\right\rangle \right\vert ^{2},
\end{equation}
where $\langle 1 | G(l) |l \rangle $ denotes the matrix elements of the resolvent
$(E-H_l)^{-1}$ between the site states in the first and $l$th slices. Given a
configuration of random potentials, $\lambda$ can be evaluated by solving Eq. (2) with
the Green's Function approach introduced in Ref. \cite{16}.

\begin{figure}[!ht]
\centering \resizebox{12cm}{!} {\includegraphics{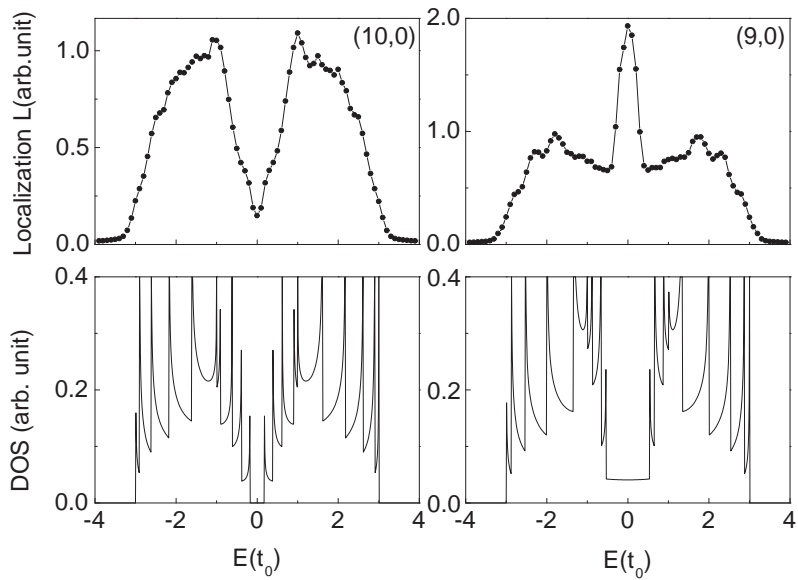}} \caption{The localization
length and the DOS without disorder as a function of energy for the $(10,0)$ and $(9,0)$
carbon nanotubes. The random potential is uniformly distributed within $[-1,1]$. }
\label{Fig-zizag}
\end{figure}

Figure 2 shows the localization length as a function of energy $E$ for
two zigzag nanotubes. The DOS for the corresponding systems without
impurities is also shown. For the semiconducting (10,0) zigzag
nanotube, $\lambda$ shows a dip at zero energy. This suppression of
$\lambda$ is apparently due to the vanishing DOS at $E=0$ since there
is a finite energy gap for the $(10,0)$ nanotube at half filling.
However, for the (9,0) zigzag nanotube, the DOS is finite and
$\lambda$ shows a peak at $E=0$. This shows that the localization
length behaves very differently in different type of nanotubes.

The appearance of the zero energy peak of $\lambda$ is a unique property of metallic
carbon nanotubes. As shown later, it does not appear in square lattice tubes. For the
metallic nanotubes, it is known that both the DOS and the energy separation between the
two DOS peaks below and above $E=0$ decrease linearly with increasing tube diameters.
However, as shown in Fig. $3$, the $E=0$ peak of $\lambda$ is found to increase
monotonically with increasing tube diameters. This indicates that the peak of $\lambda$
is anti-correlated with the DOS at $E=0$ for the metallic nanotubes.

In both $(9,0)$ and $(10,0)$ nanotubes, the DOS without impurities
shows many diverging peaks at band edges. However, these
singularities are smeared out by disorders and $\lambda$ only
exhibits some small fluctuation at the energies corresponding to
these singularities (Fig. $2$).

\begin{figure}[!ht]
\centering \resizebox{12cm}{!} {\includegraphics{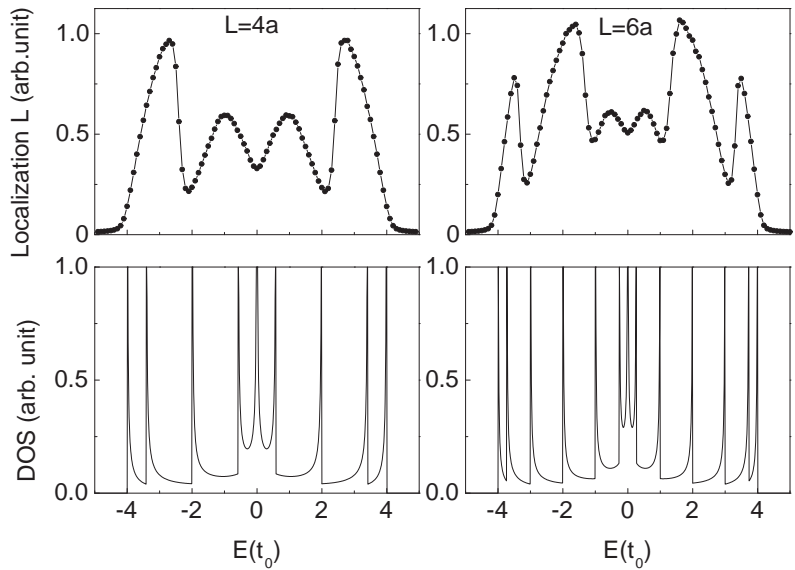}} \caption{The localization
length and the DOS without impurities as a function of energy for square lattice
nanotubes with the circumference of 4a and 6a (a is the lattice constant), respectively.
The random potential is uniformly distributed within $[-1,1]$.} \label{Fig-square}
\end{figure}

The difference of $\lambda$ in the semiconducting and metallic
nanotubes can be understood from the difference in their DOS. In a
disorder system, within the relaxation time approximation, the mean
free path $l_{e}$ is related to the DOS by the relation, \cite{17}
\begin{equation}
l_{e}\simeq \frac{\hbar v_{F}}{2\pi W^2\rho (\omega _{F})}
\end{equation}
where $v_{F}$ is the Fermi velocity and $\rho (\omega _{F})$ is the
DOS at the Fermi level. In the weak scattering limit, the velocity of
carriers decreases rapidly when the Fermi level shift to the edge of
forbidden band, but the DOS tends to diverge because of the presence
of van Hove singularities. Thus the mean free path is very short.
However, for metallic nanotubes there is a flat band around $E=0$ and
the DOS is small but non-zero, the corresponding mean free path
should be much larger than the semiconducting case. Since the
localization length is roughly proportional to the mean free path $
\lambda \propto N_{C}l_{e}$, where $N_{C}$ is the number of channels
\cite{11}, it is expected the localization length of metallic
nanotubes to be much longer than that of semiconducting one in the
middle of the band. In additional, the DOS near the Fermi level
depend only on the diameter of a metallic nanotube, independent on
its chirality. The localization length of a chiral nanotube behaves
similarly as for a zigzag or armchair nanotube with close
diameter.\cite{4} Thus the above conclusion can be also applied to a
chiral nanotube.

\begin{figure}[!ht]
\centering \resizebox{12cm}{!} {\includegraphics{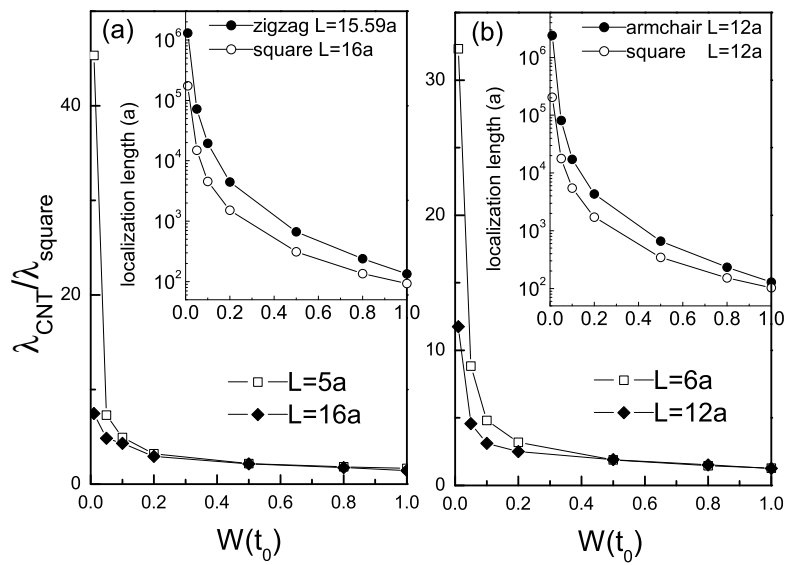}} \caption{The ratio between
the localization length of a metallic carbon nanotube and that of a square lattice
nanotube with similar circumference at $E=0$ as a function of the range of the random
potential $W$. Left panel: the circumferences of the $(9,0)$ and $(3,0)$ zigzag nanotubes are
$15.59a$ and $5.20a$, respectively. The corresponding circumferences of the square lattice
nanotubes used for comparison are $16a$ and $5a$, respectively. The inset compares the
localization length for the $(9,0)$ zigzag tube with that of the square lattice tube with
the circumference of $16a$. Right panel: the circumferences of the $(4,4)$ and $(2,2)$ armchair
nanotubes are $12a$ and $6a$, respectively. The square lattice nanotubes used for comparison
have the same circumferences. The inset compares the localization length for the $(4,4)$
armchair tube with that of the square lattice tube with the same circumference.}
\label{Fig-ratio}
\end{figure}

For comparison, we have also evaluated the localization length and
the corresponding DOS of pure systems for the square lattice
nanotubes. As shown in Fig. $3$, the DOS in the square lattice tubes
diverges at $E=0$. The impurity scattering will suppress the
divergence of DOS. However, at the Fermi level, the DOS is still very
high. Since the mean free path is inversely proportional to the DOS,
$l_e$  and $\lambda$ of square lattice nanotube are expected to be
much shorter than that of carbon nanotube as shown in Fig. 
\ref{Fig-ratio}. Unlike the metallic carbon nanotubes, $\lambda$ in a
square lattice nanotube shows a dip around $E=0$. This suggests that
$\lambda$ is strongly affected by lattice structures and the
hexagonal symmetry of carbon nanotubes can enhance the mobility of
the state at the middle of the band.

To further elucidate the difference between carbon and square-lattice nanotubes, we
evaluate the ratio between the localization length of metallic carbon nanotubes and that
of square nanotubes with close circumference. As shown in Fig. $4$, the difference between
the localization lengths in these two systems is more apparent when the circumference of
nanotubes or the strength of disorder becomes smaller. The localization length of the
(2,2) armchair nanotube (Fig. 4b) is more than a order of magnitude larger than the
square-lattice one when $W<0.5t$. This agrees with the result of White and Todorov.
\cite{11} It indicates that the mobility of carbon nanotubes is much better than that of
square lattice ones, especially for thin tubes.

In summary, we have evaluated the localization length and DOS of carbon nanotubes
and they are compared with the corresponding quantities for
square lattice nanotubes. Our results indicates that in the same
strength of disorders, electrons are more extended in a metallic
carbon nanotube than in a nanotube rolled from a square lattice,
especially in the limit of small diameters. Thus high conductance
is favored by the hexagonal structure of carbon nanotubes.
\\

This work has been supported by the National Natural Science Foundation of China under grant No $90203006$.

\end{document}